
\baselineskip 13pt
\parskip 2pt

\footline{\hfill---\enspace{\oldstyle\the\pageno}\enspace---\hfill}

\hsize=145 truemm 
\vsize=235 truemm 

\hoffset=7 truemm  
\voffset=7 truemm 

\ifx\pdfoutput\undefined
 
  \else\pdfpagewidth=210 true mm
        \pdfpageheight=297 true mm
    \fi

\def\section #1. #2. {\bigskip\noindent{\bf#1.\enspace#2.\enspace}}

\def\ie{\hbox{\it i.e.}~}
\def\bull{\raise1pt\hbox{$\scriptscriptstyle\circ$}}
\def\implies{\Rightarrow}

\def\be{$$\vbox\bgroup\halign\bgroup&$##$\hfil\cr}
\def\ee{\egroup\egroup$$}

\def\QED{\vrule width 3pt height 9pt depth -3pt}
\def\qed{\nobreak\leavevmode\hfill\QED\hskip2cm\null\smallbreak}


\font\ss=cmss8
\def\choco{{\ss CHOCO}}
\font\sevensl=cmsl10 at 7pt

\def\k{{\bf k}}
\def\IF#1\THEN#2\ELSE#3\FI{{\tt if}\ #1\ {\tt then}\ #2\ {\tt else}\ #3}
\def\T{{\tt T}}
\def\F{{\tt F}}

\def\bib[#1]#2\par{\bgroup\parindent20pt\item{[#1]}#2\par\egroup}
\def\finiteness{1}
\def\difflamb{2}
\def\christine{3}
\def\lionel{4}
\def\lionelbis{5}


\centerline{\bf A Completeness Theorem for ``Total Boolean Functions''}

\smallbreak
\centerline{July 2008}

\smallbreak
\centerline{Pierre Hyvernat,\footnote*{Laboratoire de
Math\'ematiques, Universit\'e de Savoie, 73376 Le Bourget-du-Lac Cedex,
France. This work has been partially funded by the French project
\choco\ ({\tt ANR-07-BLAN-0324}).}
{\tt pierre.hyvernat@univ-savoie.fr}}

\bigbreak
\bgroup\narrower\noindent
{\bf Abstract.} In [\christine], Christine Tasson introduces an algebraic
notion of totality for a denotational model of linear logic.  The notion of
total boolean function is, in a way, quite intuitive. This note provides a
positive answer to the question of completeness of the ``boolean centroidal
calculus'' w.r.t. total boolean functions.
\par
\egroup
\bigbreak



\section 0. Introduction. 
Even though the question answered in this note has its roots in denotational
semantics for the differential $\lambda$-calculus ([\difflamb]
and~[\finiteness], see also~[\lionel]), no background in proof-theory is
necessary to understand the problem. In the end, it boils down to a question
about a special kind of polynomials in~$2n$~variables over an arbitrary
field~$\k$. This note is almost ``self-contained'', assuming only mild
knowledge about polynomials and vector spaces (and a modicum about affine
spaces).

The only exotic (??) technology is the following formula for counting
monomials or multisets. The number of different monomials of degree~$d$
over~$n$~variables is usually denoted~$\left(\!\!\left( n \atop
d\right)\!\!\right)$. A simple counting argument shows that the number of
monomials of degree {\sl at most~$d$} in~$n$~variables is~$\left(\!\!\left(
n+1 \atop d\right)\!\!\right)$.  A closed formula for~$\left(\!\!\left( n
\atop d\right)\!\!\right)$ in terms of the usual binomial coefficient is given
by:
$$
  \left(\!\!\left( n \atop d\right)\!\!\right)
    \quad = \quad
  {n+d-1 \choose n}
  \ \hbox{.}
$$
Thus, the number of monomials of degree at most~$d$ in~$n$~variables is given
by~$n+d \choose n$.


\section 1. Total boolean polynomials. 
The category of finite dimensional vector spaces give a denotational model for
multiplicative additive linear logic. Adding the exponential is a non-trivial
task and requires infinite dimensional spaces and thus, topology. Moreover, we
need to find a subclass of spaces satisfying~$E\simeq E^{**}$. Finiteness
spaces (see~[\finiteness]) give a solution.  We won't need the details of this
technology, but it is interesting to note that objects are topological vector
spaces, and that morphisms (in the co-Kleisli category of the {\bf!}-comonad)
are ``analytic functions'', \ie power series.

Of particular interest is the space ${\bf B}$ used to interpret the booleans:
this is the vector space~$\k^2$, where~$\k$ is the ambient field.  A morphism
from~${\bf B}^n$ to~$\bf B$ is a pair~$(P_1,P_2)$ of ``finite'' power series
(polynomials) in~$2n$ variables, where each pair~$(X_{2i-1},X_{2i})$ of
variables corresponds to the $i$-th argument of the function.

\medbreak
A boolean value $(a,b)$ is {\sl total} if $a+b=1$; and a pair of polynomials is
{\sl total} if it sends total values to total values.  This means that a
pair~$(P_1,P_2)$ of polynomials in~$2n$ variables is total iff
$$
  a_1+a_2=1,\ \dots,\ a_{2n-1}+a_{2n}=1
    \quad \implies \quad
  P_1(a_1,\dots,a_{2n}) + P_2(a_1,\dots,a_{2n}) = 1 \ \hbox{.}
$$
We first restrict our attention to the case of an infinite field~$\k$:
the above condition is then equivalent to the stronger condition (a pair of
polynomials satisfying this condition is called {\sl strongly total})
$$
  P_1\big(X_1,1-X_1,\dots,X_{2n-1},1-X_{2n-1}\big)
  \ +\ %
  P_2\big(X_1,1-X_1,\dots,X_{2n-1},1-X_{2n-1}\big) = 1 \ \hbox{.} \leqno{(*)}
$$
The proof of this is easy but interesting: refer to any algebra textbook
(``Algebra'' by Lang, corollary~1.7 in chapter~IV for example) if you are in a
hurry...

\medbreak
The constructions presented below also work for finite fields, but give a
weaker result: see the remark at the end of section~5.

\proclaim Lemma.
\item\bull Strongly total polynomials form an affine subspace
of~$\k[X_1,\dots,X_{2n}]\times\k[X_1,\dots,X_{2n}]$;
\item\bull total polynomials form an affine subspace
of~$\k[X_1,\dots,X_{2n}]\times\k[X_1,\dots,X_{2n}]$.


\section 2. The centroidal calculus for boolean functions.  
The centroidal calculus produces pairs of polynomials $(P_1,P_2)$ using
\item\bull constants: $\T:=(1,0)$ and $\F:=(0,1)$;
\item\bull pairs of variables: $(X_1,X_2)$;
\item\bull $\IF (P_1,P_2) \THEN (Q_1,Q_2)\ELSE (R_1,R_2)\FI :=
  (P_1 Q_1 + P_2 R_1\,,\,P_1 Q_2 + P_2 R_2)$;
\item\bull affine combinations: $\sum_{i=1}^n \alpha_i (P_{i,1},P_{i,2})$
  where~$\sum_{i=1}^n \alpha_i = 1$.

\noindent
A pair of polynomials is {\sl centroidal} if it is generated by the above
operations.
\proclaim Lemma.
Centroidal polynomials form an affine subspace
of~$\k[X_1,\dots,X_{2n}]\times\k[X_1,\dots,X_{2n}]$.

\smallbreak
\noindent
{\sl A note on terminology:} ``affine calculus'' would be a much better name
than ``centroidal calculus''; but in the context of linear logic, this would
lead to endless confusion.

\bigbreak
The following proposition answers the natural question that was raised by
Christine Tasson and Thomas Ehrhard:
\proclaim Proposition.
Suppose the field $\k$ is infinite; then the spaces of centroidal polynomials
and of total polynomials coincide.

That centroidal polynomials are total is a direct consequence of their
definition: all centroidal polynomials are in fact {\sl strongly total}, in
the sense of~$(*)$. The rest of this note is devoted to the converse.


\section 3. Tips and tricks for centroidal polynomials.  
Here is a collection of recipes for constructing centroidal polynomials:
\item\bull $(\alpha,1-\alpha) := \alpha\,\T + (1-\alpha)\,\F$;
\item\bull $\lnot (P_1,P_2) = (P_2,P_1) := \IF (P_1,P_2)\THEN \F\ELSE \T\FI$;
\item\bull $(P_1,P_2)*(Q_1 ,Q_2) = (P_1 Q_1\,,\,P_1 Q_2 + P_2) :=
  \IF (P_1,P_2)\THEN (Q_1,Q_2) \ELSE\F\FI$;\footnote*{This operation is
  neither commutative nor associative!}
\item\bull $(P_1,P_2)^+ = (P_1+P_2\,,\,0) := \IF (P_1,P_2)\THEN\T\ELSE\T\FI$;
\item\bull $\pi_1(P_1,P_2) = (P_1\,,\,1-P_1) := \F + (P_1,P_2)^+ - \lnot
  (P_1,P_2)$.

\medbreak
\noindent
Using those, we can get more complex centroidal polynomials:

\smallbreak
\item{\sevensl(a)} suppose $P_1$ is {\sl any polynomial}; we can always get a
  centroidal term~$(P_1,P_2)$ for some polynomial $P_2$:
  \itemitem- using ``$\_*\_$'', we can get any monomial~$(M,\dots)$,
  \itemitem- if $M$ is such a monomial, $\alpha$ its coefficient in~$P_1$
    and~$m$ the total number of monomials in~$P_1$, $(m \alpha M,\dots) =
    \IF(m \alpha,1-m \alpha)\THEN(M,\dots)\ELSE\F\FI$,
  \itemitem- we can then sum those monomials using coefficients $1/m$ to get
    $(P_1,\dots)$.

\smallbreak
\item{\sevensl(b)} If $(P_1,0)$ is centroidal and if $Q_1$ is {\sl any
  polynomial}, then $\big((P_1-1) Q_1,1\big)$ is centroidal:
  \itemitem- thanks to the previous point, we can obtain $(Q_1,Q_2)$ for some $Q_2$,
  \itemitem- $\big((P_1-1)Q_1,1\big) = \big((Q_1,Q_2)*(P_1,0)\big) + \F -
    (Q_1,Q_2)$.

\smallbreak
\item{\sevensl(c)} If $(P_1,P_2)$ is centroidal and if $Q_1$ is {\sl any
  polynomial}, then $(P_1+Q_1\,,\,P_2-Q_1)$ is also centroidal:
  $(P_1+Q_1\,,\,P_2-Q_1) = (P_1,P_2) + (Q_1+Q_2,0) - (Q_2,Q_1)$

\medbreak
The last point implies in particular that it is equivalent to show
that~$(P_1,P_2)$ is centroidal and to show that~$(P_1+P_2,0)$ is centroidal.


\section 4. An interesting vector space.  
Write $\k[X_1,\dots,X_n]_{d}$ for the vector space of polynomials of degree at
most~$d$. The operator~$\varphi:\k[X_1,\dots,X_{2n}]_{d} \to
\k[X_1,\dots,X_n]_{d}$ with
$$
  \varphi\quad :\quad
  P(X_1,\dots,X_{2n}) \ \mapsto\  P(X_1,1-X_1,\dots,X_n,1-X_n)
$$
is linear and surjective.  Since the dimension of $\k[X_1,\dots,X_n]_{d}$ is
$n+d\choose n$, we get
$$
  \dim\big(\ker(\varphi)\big)
    \quad =\quad
  {2n+d \choose 2n} - {n+d\choose n}
  \ \hbox{.}
$$

\bigbreak
It is easy to see that the following polynomials are all in the kernel of $\varphi$:
$$
  \bigg((X_1+X_2)^{i_1} \times \dots \times (X_{2n-1}+X_{2n})^{i_n} - 1\bigg)
  \times X_1^{j_1}\times \dots \times X_{2n-1}^{j_n}
$$
where $(\sum_k i_k)+(\sum_k j_k)\leq d$ and at least one of the $i_k$ is non
zero.

\proclaim Lemma.
The above polynomials are linearly independent.

\smallbreak
\noindent
{\sl Proof: } suppose $\sum \alpha_k P_k = 0$ where each $P_k$ is one of the
above vectors. We show that the coefficient of
any~$\big((X_1+X_2)^{i_1}\dots(X_{2n-1}+X_{2n})^{i_n} - 1\big) X_1^{j_1}\dots
X_{2n-1}^{j_n}$ is zero by induction on~$\sum_k j_k$.

\smallbreak
\item\bull If $\sum_k j_k=0$: since the linear combination is zero, this
  implies that the global coefficient of each monomial is zero. Since
  $(X_1+X_2)^{i_1}\dots(X_{2n-1}+X_{2n})^{i_n} - 1$ is the only polynomial
  contributing to the monomial $X_2^{i_1}\dots X_{2n}^{i_n}$, its coefficient
  must be zero.

\smallbreak
\item\bull The polynomial $\big((X_1+X_2)^{i_1}\dots(X_{2n-1}+X_{2n})^{i_n} -
  1\big) X_1^{j_1}\dots X_{2n-1}^{j_n}$ is the only polynomial contributing to
  $X_2^{i_1}\dots X_{2n}^{i_n}X_1^{j_1}\dots X_{2n-1}^{j_n}$ because, by
  induction hypothesis, all the polynomials with fewer $X_{2k-1}$'s have zero
  for coefficient. This implies that the above coefficient is also zero...

\qed

\medbreak
\proclaim Corollary.
The above polynomials form a basis for $\ker(\varphi)$.

\smallbreak
\noindent
{\sl Proof: } the number of those polynomials is exactly ${2n+d \choose 2n} -
{n+d\choose n}$:
\item\bull the first term accounts for the polynomials with $(\sum_k
  i_k)+(\sum_k j_k)\leq d$,
\item\bull the second term removes the polynomials where all the $i_k$'s
  are zero.

\noindent
We have a family of ${2n+d \choose 2n} - {n+d\choose n}$ linearly independent
polynomials in a space of the same dimension: they necessarily form a basis.

\qed


\section 5. Back to total polynomials.  
Abusing our terminology, we say that a single polynomial~$P$ is total [resp.
centroidal] if the pair~$(P,0)$ is total [resp. centroidal].

We saw in section~3 that it is sufficient to show that all the total~$P$ are
centroidal.  Since the space of total polynomials is just the affine
space~$1+\ker(\varphi)$, the following polynomials form a basis for the space
of total polynomials:
$$
  1\ +\ \Big((X_1+X_2)^{i_1} \times \dots \times (X_{2n-1}+X_{2n})^{i_n} - 1\Big)
  \times X_1^{j_1}\times \dots \times X_{2n-1}^{j_n}
$$
We thus only need to show that each element in this basis is indeed centroidal.

Each $(X_1+X_2,0)$ is centroidal, so that each $(X_1+X_2)^{i_1}\dots
(X_{2n-1}+X_{2n})^{i_n}$ is also centroidal (using the ``$\_*\_$'' operation);
we can find a centroidal $\big(X_1^{j_1}\dots X_{2n-1}^{j_n}\,,\,Q\big)$ and
apply point~{\sl(b)} of section~3 to obtain
$$
  \Big(\big((X_1+X_2)^{i_1} \dots (X_{2n-1}+X_{2n})^{i_n} - 1\big)
   X_1^{j_1} \dots X_{2n-1}^{j_n}\ ,\ 1\Big)
$$
The ``$\_^+$'' operation allows to conclude the proof of the proposition.

\bigbreak
Everything we've done so far also apply to finite fields, but the result we
obtain is
\proclaim Proposition.
Suppose the field~$\k$ is finite; then the space of centroidal polynomials is
exactly the space of ``strongly total'' polynomials (see~$(*)$ in section~1).
This space is a strict subspace of the space of total polynomials.

\smallbreak
\noindent
{\sl Proof:} we only need to show that centroidal polynomials are a strict
subspace of total polynomials. Take the polynomial~$1+X(X+1)(X+2)\dots(X+l)$
where~$l+1$ is the cardinality of the field. This polynomial is total but not
strongly total: it thus can't be encoded in the centroidal calculus.

\qed


\section 6. Some examples: the ``parallel'' or and Gustave's function.  
In order to write smaller formulas, we occasionally use a single letter~$P$ to
denote a pair~$(P_1,P_2)$ of polynomials.

\medbreak
Using the usual encoding with the ``{\tt if}'' primitive, the usual ``or''
function is easily programmed in the centroidal calculus:
$$
  P \lor Q \quad := \quad
  \IF P\THEN\T\ELSE Q\FI = \big(P_1 + P_2 Q_1 \ ,\  P_2 Q_2\big)
  \ \hbox{.}
$$
However, this function is not commutative: in general,
$(P_1,P_2)\lor(Q_1,Q_2)$ is not the same as~$(Q_1,Q_2)\lor(P_1,P_2)$, except
for total values. To get a commutative version, one needs to use sums:
$$
  P\lor Q
    \quad := \quad
  {1\over 2}\  \IF P \THEN \T \ELSE Q\FI
  \ +\ %
  {1\over 2}\  \IF Q \THEN \T \ELSE P\FI
  \  \hbox{.}
$$
This ``or'' is indeed commutative, and $\F$ is neutral; but we do not have
$(P_1,P_2)\lor\T = \T$.

\smallbreak
The simplest really well-behaved ``or'' function seems to be the following:
\be
  P\lor Q    &\quad := \quad &
         & \IF P \THEN \T \ELSE Q\FI \cr
  &&{}+{}& \IF Q \THEN \T \ELSE P\FI \cr
  &&{}-{}& \IF P \THEN \big(\IF Q \THEN \T \ELSE \T\FI\big) \ELSE Q\FI \cr
             &\quad = \quad &
  \span \big(P_1 + Q_1 - P_1 Q_1 \ ,\ P_2 Q_2\big) \hskip100pt\cr
\ee
This ``or'' function is truly commutative, has~$\F$ as a neutral element and~$\T$ as
an absorbent element. It is probably the closest one can get to the
``parallel-or''.

\smallbreak
\noindent
{\sl Exercise:} with the above ``or'', we
have~$\big(1/2\,,1/2\big)\lor\big(1/2\,,1/2\big)=\big(3/4\,,1/4\big)$.
Design two other ``or'' functions which are truly commutative, have~$(1,0)$
for absorbent element and~$(0,1)$ for neutral element and are such that:
\item\bull $\big(1/2\,,1/2\big)\lor_1\big(1/2\,,1/2\big) = (1,0)$,
\item\bull $\big(1/2\,,1/2\big)\lor_2\big(1/2\,,1/2\big) = (0,1)$.

\bigbreak
Gerard Berry's ``Gustave function'' is a ternary boolean function.
It is the first and simplest example of stable but non-sequential function;
and it can be shown to have polynomial
$$
  G(X_1,X_2,Y_1,Y_2,Z_1,Z_2)
  \quad=\quad
  \big(X_1 Y_2 + Y_1 Z_2 + Z_1 X_2\ ,\ X_1 Y_1 Z_1 + X_2 Y_2 Z_2\big)
$$
in Lefschetz totality spaces. It is trivial matter to check that this function
is total. Here is one way to obtain it in the centroidal calculus:
\item\bull $P := (X*Y)*Z$;
  \hfill$\scriptstyle =\ \big(X_1 Y_1 Z_1 \ ,\  X_2 + X_1 Y_2 + X_1 Y_1 Z_2\big)$
\item\bull $Q := (\lnot X*\lnot Z)*\lnot Y$;
  \hfill$\scriptstyle =\ \big(X_2 Y_2 Z_2 \ ,\  X_1 + X_2 Z_1 + X_2 Y_1 Z_2\big)$
\item\bull $R := \pi_1(Y*\lnot Z)$;
  \hfill$\scriptstyle=\ \big(Y_1 Z_2 \ ,\  1 - Y_1 Z_2\big)$
\item\bull $S := \pi_1(X) + \lnot X - \T$;
  \hfill$\scriptstyle=\ \big(X_1 + X_2  - 1 \ ,\  1\big)$
\item\bull $U := \IF R \THEN S \ELSE \T\FI$;
  \hfill$\scriptstyle=\ \big(Z_2 Y_1 (X_1+X_2-1) \ ,\  1\big)$
\item\bull $V := \IF U \THEN \T \ELSE \lnot(X^+)\FI$;
  \hfill$\scriptstyle=\ \big(Z_2 Y_1 (X_1+X_2-1) \ ,\  X_1+X_2\big)$
\item\bull $H := P + Q - \lnot(V^+)$;
  \hfill$\scriptstyle=\ \big(X_1 Y_1 Z_1 + X_2 Y_2 Z_2\ ,\  X_1 Y_2 + Y_1 Z_2 + Z_1 X_2\big)$
\item\bull $G := \lnot H$.

\medbreak
Expressing the corresponding polynomial in the basis given in section~5 seems
to yield an even bigger centroidal expression:
\be
  X_1 Y_2 + Y_1 Z_2 + Z_1 X_2 + X_1 Y_1 Z_1 + X_2 Y_2 Z_2
    & \quad = \quad &
       & \Big((X_1+X_2)(Y_1+Y_2)(Z_1+Z_2)\Big) \cr
&&{}-{}& \Big(\big((X_1+X_2)(Y_1+Y_2) - 1\big)Z_1 +1\Big)\cr
&&{}-{}& \Big(\big((X_1+X_2)(Z_1+Z_2) - 1\big)Y_1 +1\Big)\cr
&&{}-{}& \Big(\big((Y_1+Y_2)(Z_1+Z_2) - 1\big)X_1 +1\Big)\cr
&&{}+{}& \Big(\big((X_1+X_2) - 1\big)Z_1 +1\Big)\cr
&&{}+{}& \Big(\big((Y_1+Y_2) - 1\big)X_1 +1\Big)\cr
&&{}+{}& \Big(\big((Z_1+Z_2) - 1\big)Y_1 +1\Big)\cr
\ee
where each basic polynomial can be expressed in the centroidal calculus using
the recipes from section~3.



\bigbreak
\centerline{\bf References}
\medskip

\bib
[\finiteness] Thomas Ehrhard, ``Finiteness Spaces''. {\sl Mathematical
Structures in Computer Science}, 15(04):615­646, 2005.

\medbreak
\bib
[\difflamb] Thomas Ehrhard and Laurent Regnier, ``The Differential
$\lambda$-calculus''. {\sl Theoretical Computer Science}, 309:1­41, 2003.


\medbreak
\bib
[\christine] Christine Tasson, ``Totality in an Algebraic Setting''.
Unpublished, 2008.

\medbreak
\bib
[\lionel] Lionel Vaux,  ``On Linear Combinations of $\lambda$-Terms''. {\sl
Lecture Notes in Computer Science, Term Rewriting and Applications}, 4533:374,
2007. See also~[\lionelbis]

\medbreak
\bib
[\lionelbis] Lionel Vaux, ``Algebraic $\lambda$-calculus''. Submitted, 2008.


\bye